\documentclass[a4paper,aps,prd,twocolumn,preprintnumbers,showpacs,superscriptaddress,nofootinbib]{revtex4}
\usepackage{graphics,graphicx,epsfig}
\usepackage{amsfonts,amsmath,amssymb}
\begin{document}

\title{New type of phase transition in gravitational theories}
\author{Xi\'an O. Camanho}
\affiliation{Department of Particle Physics and IGFAE, University of Santiago de Compostela, E-15782 Santiago de Compostela, Spain}
\author{Jos\'e D. Edelstein}
\affiliation{Department of Particle Physics and IGFAE, University of Santiago de Compostela, E-15782 Santiago de Compostela, Spain}
\affiliation{Centro de Estudios Cient\'{\i}ficos CECs, Av. Arturo Prat 514, Valdivia, Chile}
\author{Gast\'on Giribet}
\affiliation{University of Buenos Aires FCEN-UBA and IFIBA-CONICET, Ciudad Universitaria, Pabell\'on I, 1428, Buenos Aires, Argentina}
\author{Andr\'es Gomberoff}
\affiliation{Universidad Andres Bello, Departamento de Ciencias F\'{\i}sicas, Av. Rep\'ublica 252, Santiago, Chile}

\begin{abstract}
We set forth a new type of phase transition that might take place in gravitational theories whenever higher-curvature corrections are considered. It can be regarded as a sophisticated version of the Hawking-Page transition, mediated by the nucleation of a bubble in anti-de Sitter (AdS) space. The bubble hosts a black hole in its interior, and separates two spacetime regions with different effective cosmological constants. We compute the free energy of this configuration and compare it with that of thermal AdS. The result suggests that a phase transition actually occurs above certain critical temperature, ultimately changing the value of the cosmological constant. We discuss the consistency of the thermodynamic picture and its possible relevance in the context of AdS/CFT.
\end{abstract}
\maketitle

\section{Introduction}

Higher-curvature corrections to the Einstein-Hilbert (EH) action appear in any sensible theory of quantum gravity as next-to-leading orders in the effective action. Quadratic terms, for instance, such as the Lanczos-Gauss-Bonnet (LGB) action \cite{Lanczos}, appear in {\it bona fide} realizations of string theory \cite{GBstrings1,GBstrings2,GBstrings3,GBstrings4} and M-theory \cite{Mtheory1,Mtheory2}.

Interesting implications of these terms within the context of the AdS/CFT correspondence has been recently the focus of thorough investigation (see, {\it e.g.}, \cite{Brigante1,Brigante2,Buchel,Hofman,dBKP1,CE2,Paulos,CEP,dBKP2}). They allow for the holographic description of a broad family of quantum field theories (like 4d superconformal field theories with unequal central charges \cite{BMS}), as well as for the study of the fluid/gravity correspondence beyond the EH gravitational sector.

A remarkable example of higher-curvature gravity is Lovelock theory \cite{Lovelock}, the natural extension of general relativity to higher dimensions. More precisely, it is the most general theory of gravity yielding second-order field equations. General relativity and Chern-Simons gravity \cite{ZanelliR} are particular cases of Lovelock theories. Although quantum corrections are not generally of the Lovelock type, these theories provide a tractable playground that captures many important features that are rather generic. Among them, the existence of new branches of black hole solutions corresponding to vacua with different cosmological constants that pop out as soon as higher-curvature terms are brought into place.

The main motivation of the present work is that of studying transitions between the different vacua of the theory. This is important to investigate whether a new type of instability involving non-perturbative solutions occurs in the theory. Studying such transitions may uncover a mechanism that eventually leads to a built-in mechanism to reduce the (absolute) value of the effective cosmological constant.

It is a well-known fact that black holes in AdS display the so-called Hawking-Page (HP) transition \cite{HawkingPage}, {\it i.e.}, the semiclassical phase transition between thermal AdS space at finite temperature and a black hole in AdS with the same Hawking temperature. This has been further interpreted as a confinement/deconfinement phase transition in the dual CFT \cite{Witten}.

A natural question arises regarding the role of the different black hole branches and would be transitions among them. At first glance it might seem that such phase transitions are forbidden since distinct branches exhibit a different asymptotic behavior.

In this article, however, we report on a new type of phase transition taking place in higher-curvature gravities, which can be thought of as a sophisticated version of the HP transition involving different branches. We work out the simplest example given by LGB gravity --whose ordinary HP transition was studied in \cite{Cai}--, since it is enough to realize that these phase transitions occur in Lovelock theory as well \cite{comingsoon}, and possibly in a larger class of higher-curvature theories.
 
We consider the theory at finite temperature and observe that, above a critical temperature, $T_c$, the higher-curvature vacuum decays producing a bubble which hosts a black hole in its interior. This phase transition is analogous to the {\it thermalon} transition discussed in \cite{ghtw}, where the materialization of an electrically charged bubble induces the decay of a de Sitter vacuum into another with a smaller cosmological constant, but containing a black hole. In the present case, however, the bubble is not made from matter, but from the gravitational field itself. It separates two regions having a different effective cosmological constant {and ultimately changes its value throughout the whole space, thus changing the asymptotics}.

In the canonical ensemble, whether or not the transition takes place can be decided by evaluating the Euclidean action on two well-defined classical configurations. According to our free energy computation, the phase transition actually occurs, the preferable static configuration above $T_c$ being a black hole surrounded by a bubble. In Lorentzian signature we can analyze its subsequent dynamics. We see that the bubble eventually expand in such a way that it swallows up the whole spacetime in finite proper time, thus changing the asymptotics.

\subsection{Higher-curvature corrections}

The Lovelock action can be written in terms of the vierbein 1-form, $e^{a}=e_{\mu}^{a}\,dx^{\mu}$, and the curvature 2-form, $R^{ab} = d \omega^{ab}+\omega _{~c}^{a} {\scriptstyle\wedge}\, \omega^{cb}$, $\omega^{ab}_\mu$ being the (torsion-free) spin connection, 
\begin{equation}
\mathcal{I} = \sum_{k=0}^{K}{\frac{c_{k}}{d-2k}}\,\mathcal{I}_k +\mathcal{I}_{\partial} ~,
\label{LLaction}
\end{equation}
with the {\it bulk} contributions,
\begin{equation*}
\mathcal{I}_k = \int_{\mathcal{M}}\!\!\! \mathfrak{E}^{(d)}_{a_{1} \cdots a_{2k}}\, R^{a_{1}a_{2}} {\scriptstyle\wedge} \cdots {\scriptstyle\wedge}\, R^{a_{2k-1}a_{2k}} ~,
\end{equation*}
where $\mathfrak{E}^{(d)}_{a_{1} \cdots a_{k}} = \epsilon_{a_{1} \cdots a_{d}}\, e^{a_{k+1}} {\scriptstyle\wedge} \cdots {\scriptstyle\wedge}\, e^{a_{d}}$, $\epsilon_{a_{1} \cdots a_{d}}$ being the antisymmetric symbol, while $K$ is a positive integer, $K\leq \left[ (d-1)/2\right]$, and $\mathcal{I}_{\partial}$ refers to boundary terms to be discussed below. The coefficients $c_{k}$ are coupling constants with length dimensions $L^{2(k-1)}$. 

The $k^{\text{th}}$ term in the Lagrangian corresponds to the extension of the Euler characteristic in $2k$ dimensions. The zero$^{\text{th}}$ contribution is the cosmological constant term (we set $2\Lambda = -(d-1)(d-2)/L^{2}$, {\it i.e.}, $c_0 = 1/L^2$), the first term is the EH action (we normalize the Newton constant to $16\pi (d-3)!\,G_{N}=1$, {\it i.e.}, $c_1 = 1$), and the second term, quadratic in the Riemann curvature, is the LGB action (we take $c_2 = \lambda L^{2}$, and call $\lambda$ the LGB {\it coupling}).

Although Lovelock theory yields second order equations of motion, they are non-linear in the curvature. As a result, the theory admits more than one maximally symmetric solution; it has up to $K$ different (A)dS vacua with effective cosmological constants $\Lambda_{i}$, $i=1,2,\ldots,K$. They are the solutions of the $K^{\text{th}}$ order polynomial \cite{BoulwareDeser}
\begin{equation}
\Upsilon [\Lambda] \equiv \sum_{k=0}^{K}c_{k}\,\Lambda^{k} = c_{K}\prod_{i=1}^{K}\left( \Lambda -\Lambda _{i}\right) =0 ~.
\label{cc-algebraic}
\end{equation}
The theory exhibits degenerate behavior whenever two or more of these effective cosmological constants coincide. We are interested in the non-degenerate case.

~

\subsection{Lovelock black holes}

The first ingredient in our discussion is the black hole solution of the theory. Consider the ansatz 
\begin{equation}
ds^{2}=-f(r)\,dt^{2}+\frac{dr^{2}}{f(r)}+r^{2}\ d\Omega_{d-2}^{2} ~,
\label{bhansatz}
\end{equation}
where $d\Omega_{d-2}^{2}$ is the round metric on a $(d-2)$-dimensional sphere. The equations of motion reduce to a single first order differential equation for $f$, that can be easily solved in terms of $g=(1 -f)/r^{2}$ as \cite{Wheeler,Wheeler2}
\begin{equation}
\Upsilon \lbrack g]=\sum_{k=0}^{K}c_{k}\,g^{k}=\frac{\kappa }{r^{d-1}} ~,
\label{eqg}
\end{equation}
an implicit polynomial solution with up to $K$ branches where $\kappa$ is an integration constant related to the mass of the black hole \cite{Kastor},
\begin{equation}
M=\frac{(d-2)!\,\pi^{\frac{d}{2}}}{\pi\,\Gamma(\frac{d}{2})}\,\kappa =\frac{(d-2)!\,\pi^{\frac{d}{2}}}{\pi\,\Gamma(\frac{d}{2})}\,r_{H}^{d-1}\Upsilon \left[ \frac{1}{r_{H}^{2}}\right] ~,
\label{mass}
\end{equation}
$r_{H}$ being the location of the horizon. We focus on the quadratic theory which is enough to illustrate the phenomenon reported in this article. The solutions take the form \cite{BoulwareDeser}
\begin{equation}
g_{\pm}(r) = -\frac{1}{2\lambda L^{2}}\left( 1\pm \sqrt{1-4\lambda \left( 1 - \frac{\kappa }{r^{d-1}}\right)}\right) ~.
\label{GBbranches}
\end{equation}
Each of the two branches in (\ref{GBbranches}) is associated with a different value of the effective cosmological constant,
\begin{equation}
\Lambda _{\pm }=-\frac{1\pm \sqrt{1-4\lambda }}{2\lambda L^{2}} ~.
\end{equation}
Notice that $\lambda < 1/4$ is needed in order to have real non-degenerate values.

While $g_{-}$ has a well defined horizon, $g_{+}$ displays a naked singularity at the origin provided $M \neq 0$. This latter branch is also unstable \cite{BoulwareDeser}; the graviton propagator is proportional to $\Upsilon ^{\prime}[\Lambda_+]<0$, thus having the wrong sign with respect to the EH case. In the dual CFT this amounts to non-unitarity \cite{CEP}. The stable solution $g_-$ also happens to be the one that is continuously connected to the Schwarzschild-Tangherlini solution of general relativity in the $\lambda \rightarrow 0$ limit; namely
\begin{equation}
g_{-} \approx - \frac{1}{L^{2}} \left( 1 - \frac{\kappa}{r^{d-1}} \right) + \mathcal{O}(\lambda) ~.
\end{equation}
These features are also manifest for generic black holes in Lovelock theory \cite{CE}.

The existence of a second vacuum of effective cosmological constant $\Lambda_+$ in the higher-curvature theory was referred to, in \cite{BoulwareDeser}, as the theory having ``its own cosmological constant problem''. Moreover, for small LGB coupling, the curvature of such vacuum becomes very large, $\Lambda_ + \sim -1/(L^2\lambda)$. One may thereby argue that considering such a background in the quadratic theory does not make sense because higher-curvature terms cannot be neglected. However, far from removing this second vacuum, adding higher-order terms further produce a plethora of highly-curved vacua. With the purpose of understanding the implications of considering one such vacuum in the theory, we consider spaces that asymptote AdS of typical radius $(-\Lambda_+)^{-1/2}$.

\section{Generalized Hawking-Page transition}

\subsection{Boundary action}

Let us now discuss the role of the boundary terms, $\mathcal{I}_{\partial}$. Their contribution is necessary for the variational principle to be well defined. This is analogous to the Gibbons-Hawking term in general relativity \cite{GibbonsHawking}, which in the first order formalism can be written as
\begin{equation}
\mathcal{I}_{GH} = \frac{1}{d-2} \int_{\partial\mathcal{M}} \theta^{ab} \wedge \mathfrak{E}^{(d)}_{ab} ~,
\end{equation}
where $\theta^{ab}$ is the second fundamental form associated to the extrinsic curvature. Similarly, the boundary term associated to the LGB contribution reads \cite{Myers}
\begin{equation}
\mathcal{I}_{M} = \frac{2}{d-4} \int_{\partial\mathcal{M}} \theta ^{ab} {\scriptstyle\wedge}\, \left( R^{cd} - \frac{2}{3} \theta_{~e}^{c} {\scriptstyle\wedge}\, \theta^{ed} \right) \wedge \mathfrak{E}^{(d)}_{abcd} ~.
\label{Mterm}
\end{equation}
In the same way as\ bulk terms are the dimensional extension of the $2k$-dimensional Euler characteristic for closed manifolds, the corresponding boundary terms are needed in the extension of the Gauss-Bonnet theorem to manifolds with boundaries. The boundary action is then given by $\mathcal{I}_{\partial} = \mathcal{I}_{GH} + \lambda L^2\,\mathcal{I}_{M}$.

We want to explore configurations consisting of a spherical {\it bubble} dividing the spacetime in two regions, the outer being taken to asymptote AdS with effective cosmological constant $\Lambda_{+}$.
\begin{figure}[h]
\includegraphics[width=0.43\textwidth]{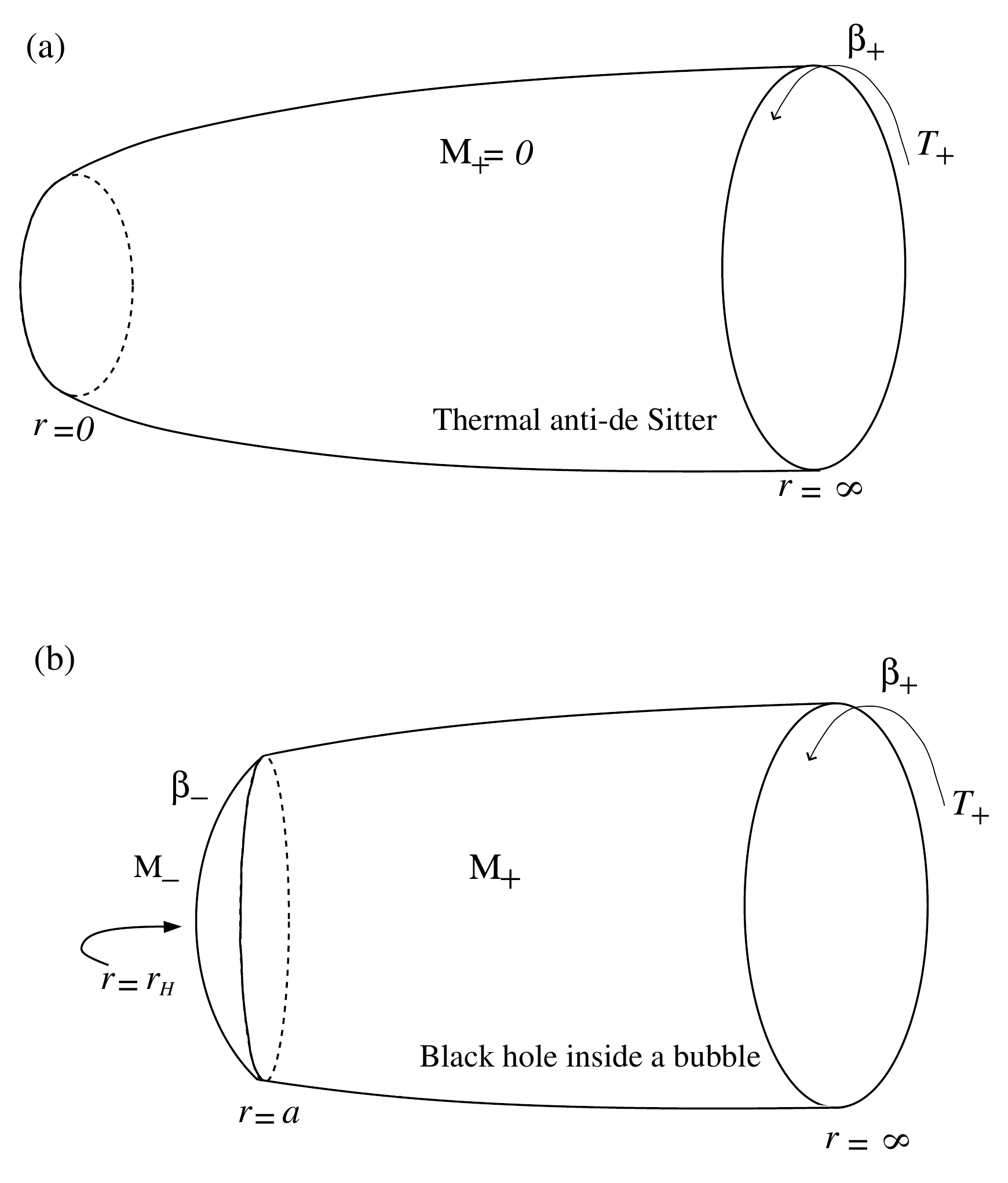}
\caption{Euclidean sections for the two possible states of the system; (a) empty thermal AdS, and (b) the bubble hosting a black hole in its interior.}
\label{CEGGFig1}
\end{figure}
Solutions consisting of a spherically symmetric surface separating regions with different vacua are known to exist \cite{wormholes, wormholes2}. They have been recently considered to explore instanton transitions, $\Lambda_+ \to \Lambda_-$, via bubble nucleation \cite{CharmPad}. In the present case we are interested in thermodynamic phase transitions. It is convenient to break the action in three pieces, 
\begin{equation}
{\mathcal{I}} = {\mathcal{I}}_{in} + {\mathcal{I}}_{\Sigma}+ {\mathcal{I}}_{out} ~.
\label{pieces}
\end{equation}
The first term is integrated inside the bubble, while the last term is integrated outside and includes all the boundary terms at infinity necessary to both have a well defined variational principle and to regularize the infinities. The term in the middle is integrated on a small region around the bubble. We consider the limit when its width goes to zero. Proceeding in this way, we deal with terms at the boundaries of each region, which give rise to a finite ${\mathcal{I}}_{\Sigma}$ in the thick-less limit \cite{Davis, GravanisWillison}. The variation of ${\mathcal{I}}_{\Sigma}$  with respect to the vierbein gives the junction conditions on the bubble \cite{Davis, GravanisWillison}. These generalize the Israel conditions of general relativity.

\subsection{The phase transition}

The configuration we will be concerned with is a {\it bubble}, whose outer region asymptotes AdS with a cosmological constant $\Lambda_{+}$, while the inner region hosts a black hole with mass $M_{-}$, and an effective cosmological constant $\Lambda_{-}$. The opposite situation does not possess a smooth Euclidean section. Across the junction, the vierbein has to be continuous. Sticking to spherical symmetry, the two bulk regions are described by a solution of the form (\ref{GBbranches}) with Euclidean signature,
\begin{equation}
ds^{2} = f_{\pm }(r)dt_{\pm}^{2} + \frac{dr^{2}}{f_{\pm }(r)}+r^{2}d\Omega_{d-2}^{2} ~,
\end{equation}
where the $\pm$ signs denote the outer/inner regions. The notation is consistent with the respective branches. We start by considering arbitrary branch solutions with undetermined mass parameters on each region. The junction conditions then fully constrain the allowed possibilities.

The junction is conveniently described by the parametric equations
\begin{equation}
r = a(\tau) ~, \quad t_{\pm} = T_{\pm}(\tau) ~,
\end{equation}
with an induced metric of the form $ds^{2} = d\tau^{2} + a(\tau)^{2} d\Omega_{d-2}^{2}$, which has to be the same from both sides. This yields
\begin{equation}
f_{\pm}(a)\,\dot{T}_{\pm }^{2} + \frac{\dot{a}^{2}}{f_{\pm }(a)} = 1 ~,
\label{RTcond}
\end{equation}
where the dot stands for derivatives with respect to the proper time $\tau$. The function $a(\tau)$ appears explicitly in the induced metric, so the radius has to be continuous across the surface. This condition allows us to write all the expressions in terms of $a$ and its derivatives and eventually find a dynamical equation for the bubble itself \cite{wormholes2}.

We are interested in static configurations which, in view of (\ref{RTcond}), translates into
\begin{equation}
\tau = \sqrt{f_{-}(a)}\,T_{-} = \sqrt{f_{+}(a)}\,T_{+} ~.
\end{equation}
This means that the physical length of the Euclidean time circle is the same as seen from both sides of the junction. This matching condition will let us determine the temperature. Once the periodicity of the inner solution is fixed by demanding regularity at the black hole horizon, that of the outer solution gets fully determined,
\begin{equation}
\sqrt{f_{-}(a)}\,\beta_{-} = \sqrt{f_{+}(a)}\,\beta_{+} ~, 
\label{Tglue}
\end{equation}
such that there is a {\it unique} free parameter, the temperature. While $\beta_{-}$ is the inverse of the Hawking temperature of the inner black hole solution, $\beta_{+}$ is the inverse of the temperature measured by an observer at infinity.

By resorting to the usual Euclidean time formalism, we can compute the free energy associated to the bubble configuration that hosts a black hole, and then compare it with that of thermal AdS at the same temperature. As we show in what follows, the computation indicates that the phase transition occurs above a critical temperature, $T_c(\lambda)$.

The canonical ensemble at temperature $1/\beta $ is defined by the path integral over all metrics which asymptote AdS identified in Euclidean time with period $\beta$,
\begin{equation}
Z = \int\! \mathcal{D}g\ {e^{-\hat{\mathcal{I}}[g]}} ~,
\label{path}
\end{equation}
where $\hat{\mathcal{I}}=-i\mathcal{I}$. The dominant contributions come from the saddle points. We have then to evaluate the Euclidean action on a classical solution,
\begin{equation}
\hat{\mathcal{I}}_{\rm cl} \simeq -\log Z =\beta F ~,
\end{equation}
which therefore gives the free energy, $F$. In the present case, this basically amounts to computing the difference between the Euclidean action of the bubble configuration and that of AdS space identified with the same period in imaginary time.

The Euclidean action is in general divergent due to the infinite volume of AdS; nevertheless, it can be suitably regularized by background subtraction, meaning that the free energy is actually measured with
respect to the maximally symmetric solution. The periodicity at infinity is fixed by demanding regularity of the black hole solution at the horizon, $r=r_H<a$, supplemented by the gluing condition (\ref{Tglue}), that determines in turn the outer periodicity. In order to simplify the discussion, we calculate the on-shell action in terms of two parameters, the position of the bubble, $a$, and the temperature, even though they are not independent from each other.

Unlike the computation of the HP effect in general relativity, here we have to consider the contribution $\hat{\mathcal{I}}_{\Sigma}$  arising on the bubble, when writing the Euclidean action in the form (\ref{pieces}). The boundary term in $\hat{\mathcal{I}}_{out}$  regularizes its divergence by subtracting the background $M_{+}=0$ with the same periodicity at infinity. The term $\hat{\mathcal{I}}_{in}$, in turn, is integrated from the horizon to the location of the bubble. Finally, $\hat{\mathcal{I}}_{\Sigma}$ is given by
\begin{equation}
\hat{\mathcal{I}}_{\Sigma} = -\hat{\mathcal{I}}_{\partial -}(a,\beta_{0}) + \hat{\mathcal{I}}_{\partial +}(a,\beta_{0}) ~,
\end{equation}
where the periodicity in Euclidean time is inherited from the bulk regions, $\beta_{0} = \sqrt{f_{\pm}(a)}\,\beta_{\pm}$. We can collect all contributions depending upon the location of the bubble, $\hat{\mathcal{I}}_{bubble}$, the rest being consequently called $\hat{\mathcal{I}}_{black\;hole}=\hat{\mathcal{I}}-\hat{\mathcal{I}}_{bubble}=\beta _{-}M_{-}-S$. 

Remarkably enough, a neat result comes out after a quite lengthy calculation --that nicely carries on to the generic Lovelock theory \cite{comingsoon}--, once the junction conditions are imposed,
\begin{equation}
\hat{\mathcal{I}}_{bubble} = \beta _{+}M_{+} - \beta _{-}M_{-} ~,
\end{equation}
which is the exact value needed to correct the on-shell bulk action such that the thermodynamic interpretation is safely preserved. In fact, due to this contribution, the total action takes the form
\begin{equation}
\hat{\mathcal{I}} = \beta_{+}M_{+} - S ~.
\label{treintaytres}
\end{equation}
That is, the bubble contributes as mass --carrying the mass difference between the two solutions-- but does not contribute to the entropy. From the Hamiltonian point of view this is naturally understood as follows. The canonical action vanishes in this case, the only possible contributions coming from boundary terms both at infinity and at the horizon, yielding respectively $\beta _{+}M_{+}$ and the entropy, which are nothing but the total charges of the solution. The junction conditions simply imply the continuity of canonical momenta \cite{Banados}.

Equation (\ref{treintaytres}) shows that the junction conditions are important to guarantee the consistency of the thermodynamic picture. They also imply 
\begin{equation}
\beta_{+}dM_{+} = \beta_{-}dM_{-} = dS ~,
\label{new1law}
\end{equation}
so that the first law of thermodynamics holds both for the whole configuration ($\beta _{+}$ and $M_{+}$) as well as for the black hole ($\beta _{-}$ and $M_{-}$).

\subsection{Bubble nucleation}

Having proven the consistency of the thermodynamic picture in the case of the bubble configuration by deriving  (\ref{treintaytres}) and (\ref{new1law}), we are ready to address the question of global thermodynamic
stability. This amounts to analyzing the free energy associated to the bubble configuration.

\begin{figure}[h]
\includegraphics[width=0.39\textwidth]{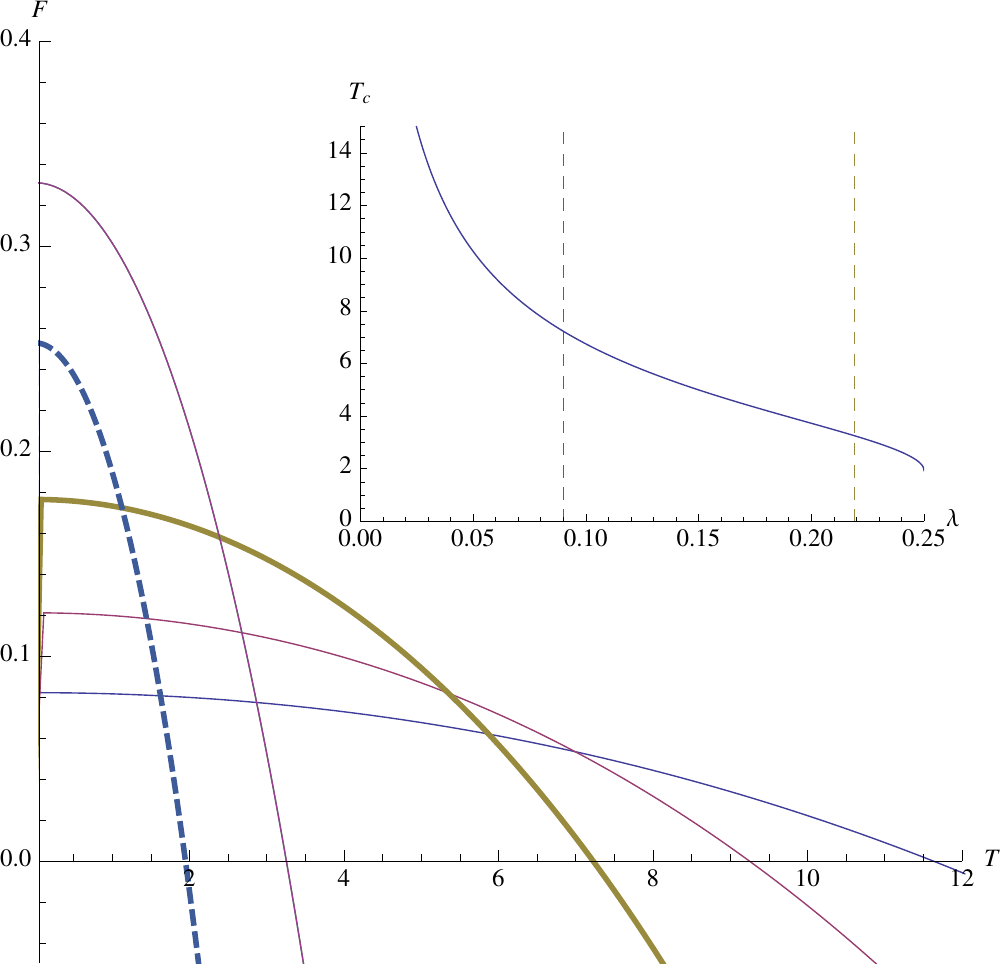}
\caption{Free energy versus temperature in 5d for $\lambda = 0.04, 0.06, 0.09$ (positivity bound), $0.219$ (maximal $F(T=0)$), and $\lambda \to 1/4$ (from right to left). The $\lambda$ dependence of the critical temperature is displayed in a separate box.}
\label{CEGGFig2}
\end{figure}

The free energy, $F$, as a function of the temperature $1/\beta_+$ displays a critical temperature above which it becomes negative and, thus, the phase transition occurs (see Fig.\ref{CEGGFig2}).
If the free energy is positive, however, the system is metastable. It decays by nucleating bubbles with a probability given, in the semiclassical approximation by $e^{-\beta_+ F}$. Therefore, after enough time, the system will alway end up in the stable, black hole solution. This is reminiscent of the HP transition, except for the fact that, here, the thermal AdS vacuum decays into a black hole {\it belonging to a different branch}. $T_c(\lambda)$ is monotonically decreasing, the phase transition becoming increasingly unlikely the more we come closer to the EH -- {\it classical} -- limit. In this sense, it is a quantum mechanical phenomenon.

These high temperature configurations, or {\it thermalons}, correspond to a black hole and a bubble in equilibrium, connecting inner and outer solutions of different branches $g_-$ and $g_+$. Such solutions exist just for positive values of $\lambda$. The infinite temperature limit corresponds to planar black holes, in which the junction conditions lead to a remarkably simple relation that is valid also for general Lovelock gravity \cite{comingsoon}, $\beta_{+} M_{+} = \beta _{-} M_{-}$. The corresponding free energy is always negative,
\begin{equation}
F = -\frac{(d-3)!\,\pi^{\frac{d}{2}}}{\pi\,\Gamma(\frac{d}{2})} \frac{r_{H}^{d-1}}{L^{2}} \frac{\beta_{-}}{\beta_{+}} = - \frac{M_{+}}{(d-2)} ~,
\end{equation}
this implying that the preferable classical solution is always the thermalon and the transition always occurs for high enough temperature. 

The junction conditions considered above determine not only the equilibrium configuration but also, in Lorentzian signature, the effective potential felt by the bubble and, consequently, its subsequent dynamics. The scalar field $a(\tau)$ specifying the location of the bubble sits at the top of a potential barrier. It will therefore either collapse or eventually expand in such a way that it engulfs the whole spacetime in finite proper time, thus changing its asymptotic behavior \cite{comingsoon}.

~

\section{Discussion}

We presented a novel mechanism for phase transitions that is a distinctive feature of higher curvature theories of gravity. These theories have several branches of asymptotically (A)dS solutions that might admit an interpretation as different phases of the dual field theory. Phase transitions among these are driven by the mechanism described in the present article. Mimicking the thermalon configuration \cite{ghtw}, a bubble separating two regions of different cosmological constants pops out, generically hosting a black hole.

This configuration is thermodynamically preferred above some critical temperature. The corresponding phase transition can be interpreted as a generalized HP transition for the high-curvature branches, driving the system towards the EH branch. From the holographic point of view this looks like a confinement-deconfinement phase transition in a dual CFT, involving an effective change in the 't Hooft coupling, both phases being strongly coupled. Whether a phenomenon like this takes place in a 4d CFT, particularly within the framework of the fluid/gravity correspondence --where both phases might be characterized by different transport coefficients--, or it is overtaken by higher curvature corrections, is an open question at this point.

The bubble configuration, being unstable, dynamically changes the asymptotic cosmological constant, transitioning towards the stable horizonful branch of solutions, the only one usually considered as relevant. This is then a natural mechanism for the system to select the general relativistic vacuum among all possible ones. We are aware of the fact that the vacuum $\Lambda_+$ in the LGB theory exhibits ghosts. The phenomenon presented in this article, however, takes place in the Lovelock theory as well, where there are further healthy vacua than the one connected to the EH action \cite{CEP}.

We think that this mechanism is quite general and deserves further investigation.

\acknowledgments
%
We wish to thank Stanley Deser, Juan Maldacena, Carlos N\'u\~nez and Josep M. Pons for their interesting comments.
G.G. thanks Ceci Garraffo, Elias Gravanis, and Steve Willison for previous collaborations in the subject.
The work of X.O.C. and J.D.E. was supported in part by MICINN and FEDER (grant FPA2011-22594), by Xunta de Galicia (Conseller\'{\i}a de Educaci\'on and grant PGIDIT10PXIB206075PR), and by the Spanish Consolider-Ingenio 2010 Programme CPAN (CSD2007-00042).
The work of G.G. was supported by NSF-CONICET, PIP, and PICT grants from CONICET and ANPCyT. 
The work of A.G. was partially supported by Fondecyt (Chile) Grant $\#$1090753.
X.O.C. and J.D.E. would like to thank the FCEN-UBA and UNAB for hospitality during part of this project.
X.O.C. is thankful to the Front of Galician-speaking Scientists for encouragement.

\end{document}